\documentclass[final]{ias2}
\usepackage{dcolumn}
\usepackage{bm}

\usepackage[utf8x]{inputenc}
\usepackage{graphicx,subfigure,setspace,color,fancyhdr,layout,appendix}
\usepackage{amsbsy,amsmath,amssymb,amsfonts}
\usepackage[numbers,sort&compress]{natbib}

\definecolor{darkgreen}{rgb}{0.0, 0.2, 0.13}
\definecolor{testgreen}{rgb}{0.11, 0.35, 0.02}
\definecolor{Navyblue}{rgb}{0.0, 0.0, 0.5}
\definecolor{magenta}{rgb}{0.79, 0.08, 0.48}
\definecolor{red}{rgb}{0.77, 0.01, 0.2}
\usepackage[colorlinks,linkcolor=blue,urlcolor=magenta,citecolor=red,plainpages=false,pdfpagelabels,breaklinks]{hyperref}

\usepackage{multirow}
\usepackage{array}
\usepackage{booktabs}
\usepackage{ctable}
\usepackage{upgreek}
\usepackage{epsfig}
\usepackage{mathrsfs}
\usepackage{amssymb}
\usepackage{amsbsy}
\usepackage{color}
\usepackage{cancel}
\usepackage{marginnote}
\newcommand{\bcen}{\begin{center}}
\newcommand{\ecen}{\end{center}}
\newcommand{\btab}{\begin{tabular}}
\newcommand{\etab}{\end{tabular}}
\newcommand{\bdes}{\begin{description}}
\newcommand{\edes}{\end{description}}

\newcommand{\beq}{\begin{equation}}
\newcommand{\eeq}{\end{equation}}
\newcommand{\bea}{\begin{eqnarray}}
\newcommand{\eea}{\end{eqnarray}}

\newcommand{\half}{\frac{1}{2}}
\newcommand{\bary}{\begin{array}}
\newcommand{\eary}{\end{array}}
\newcommand{\benum}{\begin{enumerate}}
\newcommand{\eenum}{\end{enumerate}}
\newcommand{\bitem}{\begin{itemize}}
\newcommand{\eitem}{\end{itemize}}

\newcommand{\bk} { \bm{k} }

\newcommand{\bn} { \mbox{\boldmath $n$}}

\newcommand{\bp} { \bm{p} }
\newcommand{\bq} { \bm{q} }
\newcommand{\br} { \boldsymbol{r}}

\newcommand{\dou}{\partial}

\newcommand{\D}[1]{\mbox{d}{#1}}

\newcommand{\eqn}[1] {eqn.~(\ref{#1})}

\newcommand{\fig}[1]{fig.~\ref{#1}}

\makeatletter

\newcommand{\Rmnum}[1]{\expandafter\@slowromancap\romannumeral #1@}
\makeatother

\newcommand{\myfigwidth}{0.37\paperwidth}

\newcommand{\mywidth}{0.27\paperwidth}
\newcommand{\myheight}{0.27\paperwidth}
\newcommand{\myheightcus}{0.26\paperwidth}

\newcommand{\omegazero}{{\omega_0}}

\newcommand{\mylabel}[1]{\label{#1}}

\begin{document}

\title{Spectral intensity distribution of trapped fermions}
\author[sud]{Sudeep Kumar Ghosh}
\email{sudeep@physics.iisc.ernet.in}
\address[sud]{Centre for Condensed Matter Theory, Department of Physics, Indian Institute of Science, Bangalore 560 012, India}

\begin{abstract}
To calculate static response properties of a many body system, local density approximation (LDA) can be safely applied. But applicability of LDA is limited for the case of dynamical response functions since dynamics of the system needs to be considered as well. To examine this in the context of cold atoms, we consider a system of non-interacting spin-$\half$ fermions confined by a harmonic trapping potential. We have calculated a very important response function, the spectral intensity distribution function (SIDF), both exactly and using LDA at zero temperature and compared with each other for different dimensions, trap frequencies and momenta. The behavior of the SIDF at a particular momentum can be explained by noting the behavior of the density of states (DOS) of the free system (without trap) in that particular dimension. The agreement between exact and LDA SIDFs becomes better with increase in dimensions and number of particles.
\end{abstract}


\maketitle

\section{Introduction}
\mylabel{sec:Introduction}
In recent years, cold atomic systems have been a very active area of both theoretical and experimental research because of their ability to provide clean artificial systems and wide tunability of parameters. Overcoming some initial difficulties like entropy removal, creation of gauge fields etc., these systems have emerged to be clean quantum emulators. The interaction between the particles can be tuned by changing magnetic field across a Feshbach resonance. Due to this tunability of interactions, desired many body quantum Hamiltonians can be simulated. Thus cold atomic systems allow us to study interesting quantum many body physics in a controlled environment when it is difficult to prepare natural systems with desired parameter regimes and enable us to create strongly correlated systems of interest in many areas like condensed matter, nuclear physics, high-energy physics etc.

The temperature being very low, trapped cold atomic gases are in the quantum degeneracy regime. In this regime bosons condense to a Bose Einstein Condensate (BEC).\cite{noble} By trapping the BEC in an optical potential and suppressing charge fluctuations by increasing the depth of the optical lattice, transition from a superfluid BEC to a Mott insulator had been realized.\cite{Greiner_nat2002} Similarly, in degenerate Fermi gases, crossover from a condensate of very weakly bound cooper pairs in the Bardeen-Cooper-Schrieffer (BCS) state to a condensate of tightly bound Bosons in the BEC state, was obtained by tuning the interaction between the Fermions across unitarity.\cite{Regal2006} Interesting quantum many body physics occurs in the presence of magnetic fields but since cold atoms are neutral, gauge fields are needed to be produced artificially.\cite{Lin_nat2009,Philips2009,Dalibard2011} In Fermions synthetic non-Abelian gauge fields have many interesting consequences like emergence of novel states\cite{Sudeep2011}, amplification of interaction etc.\cite{Jayantha_twobody,Hu2011,Jayantha2011} Experimentally these synthetic non-Abelian gauge fields have been realized recently.\cite{Wang2012,Cheuk2012}

Another major achievement of the last decade is development in the field of photo emission spectroscopy.\cite{Comin2013,Andrea2003} Photo-emission spectroscopy enables us to probe occupied single-particle states of a Fermionic system. The usual photo-emission spectroscopy was generalized to the case of cold atomic systems using radio frequency photo-emission spectroscopy.\cite{Dao2007} In Fermi gases this technique reveals interesting many body effects like the pseudogap phenomenon.\cite{Jin_nat2008,Jin_nat2010,Brend2011,Ville2012}

The spectral function is a very important quantity for a many body system since it reveals many important aspects like excitations of the system. In radio frequency photo-emission spectroscopy, this quantity is routinely measured.\cite{Jin_nat2010}  The static response properties of a many body system are well captured within LDA since the dynamics of the system is not involved in this case. To account for the dynamics of the system in obtaining dynamical response functions, there are different versions of LDA like the time dependent LDA (TDLDA). But their applicability depends on the adiabaticity of the dynamics of the system.\cite{Kohn1985,Yabana1999} Let $\chi (\omega^+,\bq)$ be a response function of a many body system, where $\bq$ is the momentum and $\omega$ is the frequency, then to obtain the static response function $\lim_{\omega \to 0}\chi (\omega^+,\bq)$ or the average static response $\lim_{\omega \to 0}\lim_{\bq \to 0}\chi (\omega^+,\bq)$, LDA can be applied. But in obtaining the full dynamical response function $\chi (\omega^+,\bq)$, applicability of LDA is limited because of the reason described earlier.

Even a system of non-interacting fermions in the presence of trap has many interesting behaviors since the trapping potential explicitly breaks spatial translational invariance.\cite{Sudeep2011,Stringari04,Mueller04,Gleisberg00,Vignolo00} The chemical potential and thermodynamic quantities like the specific heat show oscillations as a function of number of particles due to the finite degeneracy of the energy eigenstates of the isotropic harmonic oscillator.\cite{Butts97,Toms05,Schneider98} So, we take the viewpoint that this ideal Fermi system in the presence of an isotropic harmonic trapping potential is very interesting and consider spin-$\half$ version of it at zero temperature. We investigate the performance of LDA with respect to the exact case by calculating a dynamical response function of the system as an example. The dynamical response function, we calculate, is the spectral intensity distribution function (SIDF). This SIDF is calculated exactly and using LDA in any dimension for various parameter regimes. The results obtained using the two techniques are compared for different momenta, different trap frequencies and different number of particles in the physical dimensions $1$, $2$ and $3$. We have found that as the dimension increases the agreement between the exact and LDA SIDFs gets better. Moreover, increase in number of particles makes LDA better in higher dimensions. In the following section, we describe the system under consideration and formulate the problem.

\section{Description of the system}
\mylabel{sec:system}
We consider a system of non-interacting spin-$\half$ fermions trapped in a harmonic potential of frequency $\omegazero$. The Hamiltonian of the system is 
\beq
\hat{{\cal H}} = \frac{\hat{\bp}^2}{2 m} + \half m{\omegazero}^2 \hat{\br}^2 \,\,,
\eeq
where $m$ is the mass of fermions. Taking $m$ and the Plank's constant ($\hbar$) to be unity, the Hamiltonian operator in momentum space can be written as
\beq
\hat{{\cal H}} = -\half {\omegazero}^2 \frac{\dou^2}{\dou {\hat{\bp}}^2} + \half {\hat{\bp}}^2 \,\,.
\eeq
The one dimensional harmonic oscillator eigenstates in momentum space can be labeled by a quantum number $n$ ($n$ is an integer). They are
\beq
\phi(n,k) = \frac{1}{\sqrt{2^n n!}} \left(\frac{1}{\pi \omegazero}\right)^{\frac{1}{4}} exp\left(-\frac{k^2}{2\omegazero}\right) H_n\left(\frac{k}{\sqrt{\omegazero}}\right) \,\,,
\eeq
with energy $\varepsilon_n = (n + \half) \omegazero$. The eigenstates of the harmonic oscillator in $d$-dimension are products of $d$ one dimensional harmonic oscillator eigenfunctions and are labeled by a set of $d$ integers $\{ \bn \}$. In momentum space representation they are given by 
\beq
\langle \bk | \bn \rangle = \psi(\bn , \bk) = \prod_{i = 1}^{d} \phi(n_i , k_i) \,\,,
\label{eqn:eigenfunction}
\eeq 
with energies $\varepsilon(\bn) = \sum_{i = 1}^{d} (n_i + \half) \omegazero$. In calculating the SIDF exactly these eigenfunctions are used while in LDA the harmonic potential are used to obtain approximate local quantities.

\section{Calculation of the SIDF}
\mylabel{sec:Intensity}

The SIDF is an important response function since it contains information about many body effects in the system. At zero temperature, the SIDF is calculated in this section using two techniques: A.) LDA and B.) Exact method. Complete analytical expressions for the SIDF is obtained using LDA in any dimension and exact SIDF is obtained numerically. We note that there are two important scales in the problem: trap center chemical potential ($\mu_0$) and the frequency of the harmonic potential ($\omega_0$).

\subsection{Calculation using LDA}
Within LDA, we take into account the effect of the trap potential by making the chemical potential $\mu$ of the system a ``local variable''.\cite{Butts97} This local chemical potential $\mu(\br)$ determines the density of particles $\rho(\br)$ at a point in space $\br$ and they are related via the equation of state. As a function of $\mu_0$, $\mu(\br)$ is given by
\beq
\mu (\br) + \frac{1}{2} \omegazero^2 r^2 = \mu_0 \,\,.
\label{eqn:LDAmu} 
\eeq
The SIDF calculation within LDA is, therefore, two fold: first obtain $\mu_0$ as a function of $\omega_0$ and number of particles (N) and then calculate $\mu(\br)$ to obtain SIDF. 

\subsubsection{Calculation of the trap center chemical potential}
Consider there are $N$ particles in the trap which forms a cloud of radius $R_0$. Then in $d$-dimension ($d \geq 2$) we have
 \beq
 \int_0^{R_0} \D{\br} \; \rho(\br) = N \,\,,
 \eeq 
 where $\rho(\br)$ is the density of particles at position $\br$. Considering spherical symmetry in the problem we note that $\rho(\br)$ depends only on $r$, which leads to
 \beq
 C(d) \int_0^{R_0} \D{r} \; r^{d-1} \rho({r}) = N \,\,,
\label{eqn:radius} 
 \eeq
with $C(d) = \frac{2 \pi^{\frac{d}{2}}}{\Gamma(\frac{d}{2})}$ . The trap center chemical potential $\mu_0$ and the radius of the cloud $R_0$ is related as
\beq
\mu_0 = \half \omegazero^2 {R_0}^2 \,\,. 
\label{eqn:muzero}
\eeq
If $g(\bk)$ is the DOS of non-interacting free spin-$\half$ particles in $\bk$-space then  $g(\bk) = 2 \frac{V}{(2\pi)^d}$ (the $2$ factor comes from spin degeneracy and $V$ is volume of the $d$ spatial dimension) in $d$-dimension. So, for $d \geq 2$ 
\beq
\int_0^{k_F} \D{\bk} \; g(\bk) = N \,\,,
\eeq
where $k_F$ is the Fermi momentum of the system. The density of particles $\rho$ is therefore
\beq
\rho = \frac{2 C(d)}{d (2 \pi)^d} (2 \mu)^{\frac{d}{2}} \,\,,
\label{eqn:density}
\eeq
 where $\mu$ is the chemical potential of the free Fermi gas. This is the equation of state of the free Fermi gas in $d (\geq 2)$ dimensions. Using this equation of state, from \eqn{eqn:LDAmu} and \eqn{eqn:density} within LDA, the spatial density of the system can be written as
\beq
\rho(\br) = \frac{2 C(d)}{d (2 \pi)^d} (2 \mu_0 - \omegazero^2 {r}^2)^{\frac{d}{2}} \,\,.
\label{eqn:LDAdensity}
\eeq 

Using \eqn{eqn:LDAdensity} in \eqn{eqn:radius} and then \eqn{eqn:muzero}, we obtain a relation among the variables $\mu_0$, $\omegazero$ and $N$ in $d$-dimension ($d\geq 2$)
\beq
\left(\frac{2 \mu_0}{\omegazero}\right)^d = \frac{N}{X(d)} \,\,,
\label{eqn:dtrap}
\eeq
where $X(d) = \frac{\sqrt{\pi}}{d (2^{2d-2})\Gamma(\frac{d}{2})\Gamma(\frac{d+1}{2})}$. Similarly, for a system of $N$ particles in $1$-dimension 
\begin{eqnarray}
&&\int_{-k_F}^{k_F} \D{k} \; g(k) = N \,\,, \nonumber \\
&&\Rightarrow \sqrt{2 \mu} = \frac{\pi \rho}{2} \,\,.
\end{eqnarray}

The spatial density of the trapped particles $\rho(r)$ obtained in the same way is
\beq
\rho(r) = \frac{2 \omegazero}{\pi} \sqrt{{R_0}^2 - r^2} \,\,,
\eeq
and we have
\beq
\int_{-R_0}^{R_0} \D{r} \; \rho(r) = N \,\,.
\eeq 
The above equation then leads to 1d version of \eqn{eqn:dtrap} to be
\beq
\frac{2 \mu_0}{\omegazero} = N \,\,.
\label{eqn:1dtrap}
\eeq

Combining \eqn{eqn:dtrap} and \eqn{eqn:1dtrap}, the trap center chemical potential $\mu_0$, the trap frequency $\omega_0$ and the number of particles $N$ in any dimension $d$ are related as
\beq
 X(d) (\omegazero {R_0}^2)^d = N \,\,.
\eeq
This relation is used to calculate the local chemical potential.

\subsubsection{Obtaining the SIDF within LDA}
 The single particle spectral function of a system with dispersion relation $\epsilon(\bk)$ is 
\beq
A(\bk, \omega) = \delta(\omega-\epsilon(\bk)) \,\,.
\eeq

In LDA, the SIDF is obtained by adding local single particle spectral functions with suitable spectral weights. We approximate the free particle dispersion $\epsilon(\bk)$ to a local dispersion $\epsilon(\br,\bk)$ at every position $\br$ in space as
\beq
\epsilon(\br,\bk) = \frac{k^2}{2} + \half \omegazero^2 r^2 \,\,,
\eeq
for the trapped system. The LDA form of the local single particle spectral function $A(\br,\bk,\omega)$ is then
\beq
A(\br,\bk,\omega) = \delta(\omega - \epsilon(\br,\bk)) \,\,.
\eeq
So, the total SIDF $I(\bk,\omega)$ is obtained as
\beq
I(\bk,\omega) = \int \D{\br} \; W(\br,\bk) A(\br,\bk,\omega) \,\,,
\label{eqn:LDAspectra}
\eeq 
where $W(\br,\bk)$ is the spectral weight of the local single particle spectral function at position $\br$. In $d$-dimension it is given by
$$
W(\br,\bk) = \frac{1}{(2 \pi)^d} \Theta(\mu_0 - \epsilon(\br,\bk)) \,\,.
$$

In $1$-dimension, we use the relation between the trap center chemical potential $\mu_0$ and the number of particles $N$ from \eqn{eqn:1dtrap} in \eqn{eqn:LDAspectra} to obtain
\beq
I(k,\omega) = \frac{1}{2\pi} \int_{-r_k}^{r_k} \D{r} \; \delta(\omega-\frac{k^2}{2}-\half \omegazero^2 r^2) \Theta(\mu_0 - \omega)\,\,,
\eeq 
where $r_k = \frac{\sqrt{2 \mu_0 - k^2}}{\omegazero}$. This equation gives in $1$-dimension the LDA-form of the SIDF
\beq
I(k,\omega) = \frac{ \Theta(\mu_0 - \omega) \Theta(\omega - \frac{k^2}{2})}{\pi \omegazero \sqrt{2 \omega -k^2}} \,\,.
\label{eqn:LDA1inten}
\eeq
In the same way as above in $d$-dimension ($d\geq  2$) using \eqn{eqn:dtrap} in \eqn{eqn:LDAspectra} we get
\beq
I(\bk,\omega) = \frac{ C(d)}{(2\pi)^d} \int_0^{r_k} \D{r} \; r^{d-1} \delta(\omega-\frac{k^2}{2}-\half \omegazero^2 r^2) \Theta(\mu_0 - \omega) \,\,,
\eeq
which then gives the SIDF for spin-$\half$ particles in $d$-dimension ($d\geq  2$) to be 
\beq
I(\bk,\omega) = \frac{C(d)}{(2\pi)^d} \left(\frac{{R_m}^{d-2}}{\omegazero^2}\right) \Theta(\mu_0 - \omega) \Theta(\omega - \frac{k^2}{2}) \,\,,
\label{eqn:LDAdinten}
\eeq
where $R_m = \sqrt{\frac{2\omega - k^2}{\omegazero^2}}$.\\

The form of the SIDF $I(\bk,\omega)$ in any dimension $d$ from \eqn{eqn:LDA1inten} and \eqn{eqn:LDAdinten} can therefore be written as
\beq
I(\bk,\omega) = \frac{C(d)}{(2\pi)^d} \left(\frac{{R_m}^{d-2}}{\omegazero^2}\right) \Theta(\mu_0 - \omega) \Theta(\omega - \frac{k^2}{2}) \,\,,
\label{eqn:LDAinten}
\eeq
where $R_m = \sqrt{\frac{2\omega - k^2}{\omegazero^2}}$ and $C(d) = \frac{2 \pi^{\frac{d}{2}}}{\Gamma(\frac{d}{2})}$.

Due to the conservation of number of particles, the SIDF obeys the following sum rule
\beq
\int \D{\bk} \; \D{\omega} \; I(\bk,\omega) = N \,\,.
\label{eqn:sumrule}
\eeq
By using the expression of $I(\bk,\omega)$ from \eqn{eqn:LDAinten} and carrying out the integration over momentum and energy we see that the SIDFs obtained using LDA indeed obey the sum rule given by \eqn{eqn:sumrule}.

\subsection{Exact calculation}
To calculate the SIDF exactly we use the harmonic oscillator basis set given by \eqn{eqn:eigenfunction}. In terms of these eigenstates labeled by a set of quantum numbers $\{\bn\}$ with energies $\varepsilon({\bn})$, the SIDF is
\beq
I(\bk,\omega) = \sum_{\{ \bn \}}\; |{\langle \bn | \bk \rangle}|^2 \delta(\omega - \varepsilon({\bn})) \,\,,
\label{eqn:exactSIDF}
\eeq
where the sum is over all the occupied states. The SIDF thus obtained is spherically symmetric. Orthonormality of the harmonic oscillator eigenfunctions can be used to show that the sum rule given by \eqn{eqn:sumrule} is satisfied. By using these eigenfunctions and carrying out the summation over all the occupied states for a fixed number of particles $N$, we obtain the SIDF in any dimension. The summation is performed numerically by broadening the delta function as a Lorentzian of `suitable width'. As an approximation to the delta function $\delta(x)$, we use the Cauchy-Lorentz probability density function
\beq
{\cal L}(x,\eta) = \frac{1}{\pi} \frac{\eta}{x^2 + \eta^2} \,\,,
\eeq 
where $\eta$ is the half width at half maximum (HWHM) of ${\cal L}(x,\eta)$. As $\eta\rightarrow 0$, the Lorenzian distribution ${\cal L}(x,\eta)$ approaches the delta function distribution $\delta(x)$; where $\eta$ is a phenomenological parameter which needs to be chosen properly. Since the energy levels of an isotropic harmonic oscillator are degenerate, we account for suitable degeneracies of these levels and choose number of particles such that the highest occupied harmonic oscillator level is completely filled. The degeneracy $g_d^n$ of the $n$-th energy eigenstate of a $d$-dimensional isotropic harmonic oscillator can be calculated from the following recursion relation
\beq
g_d^n = \sum_{n_1 = 0}^{n} g_{d-1}^{n-n_1} \,\,\,\,\,\text{with} \,\,\, g_1^n = 1 \,\,.
\eeq 
Using this relation, the expression for the degeneracy is obtained to be
\beq
g_d^n = \frac{\Gamma(n+d)}{\Gamma(d) \Gamma(n+1)} \,\,,
\eeq
and if the highest occupied energy level is $n = \nu_{max}$, the total number of particles in the system (including spin degeneracy) is
\beq
N_d^{\nu_{max}} = 2 \sum_{n = 0}^{\nu_{max}} g_d^n = \frac{2 \,\Gamma(\nu_{max}+d+1)}{\Gamma(d+1) \Gamma(\nu_{max}+1)} \,\,.
\eeq
The next section contains discussions of our results.

\section{Results and discussion}
\mylabel{sec:comparison}
It is noted from \eqn{eqn:LDAinten} that the definition of the SIDF at a particular momentum has behavior similar to the DOS of the system as a function of energy without the trapping potential at that particular dimension. So, we expect the SIDF of the trapped spin-$\half$ fermionic system will have behavior similar to the corresponding DOS of the free spin-$\half$ fermionic system. The DOS, $g(\omega)$, as a function of $\omega$ of a free spin-$\half$ fermionic system in $1$, $2$ and $3$ dimensions, has behaviors $\frac{1}{\sqrt{\omega}}$, constant and $\sqrt{\omega}$ respectively. Thus $I(\bk,\omega)$ of the trapped system at a particular momentum is expected to have similar behaviors in the corresponding dimensions. The dynamical response function, $I(\bk,\omega)$ encodes the dynamics of the system and have information about the excitations of the system. To capture the dynamical nature within LDA, there are different forms of LDA like the TDLDA but their domain of applicability is limited by the slowness of dynamics of the system.\cite{Kohn1985,Yabana1999} The simplest version of LDA (meaning the time independent LDA) is used here to calculate the dynamical response function $I(\bk,\omega)$ and its behavior is studied in different parameter regimes within this approximation. The SIDF obtained using LDA is compared with corresponding exact numerical results for different values of $k = |\bk|$ due to the spherical symmetry. We discuss their behaviors only for the physical dimensions $1$, $2$ and $3$ although in principle the SIDFs can be calculated using these two methods in any dimension.

\begin{figure}
\centerline{\includegraphics[width=\myfigwidth, height=\myheight]{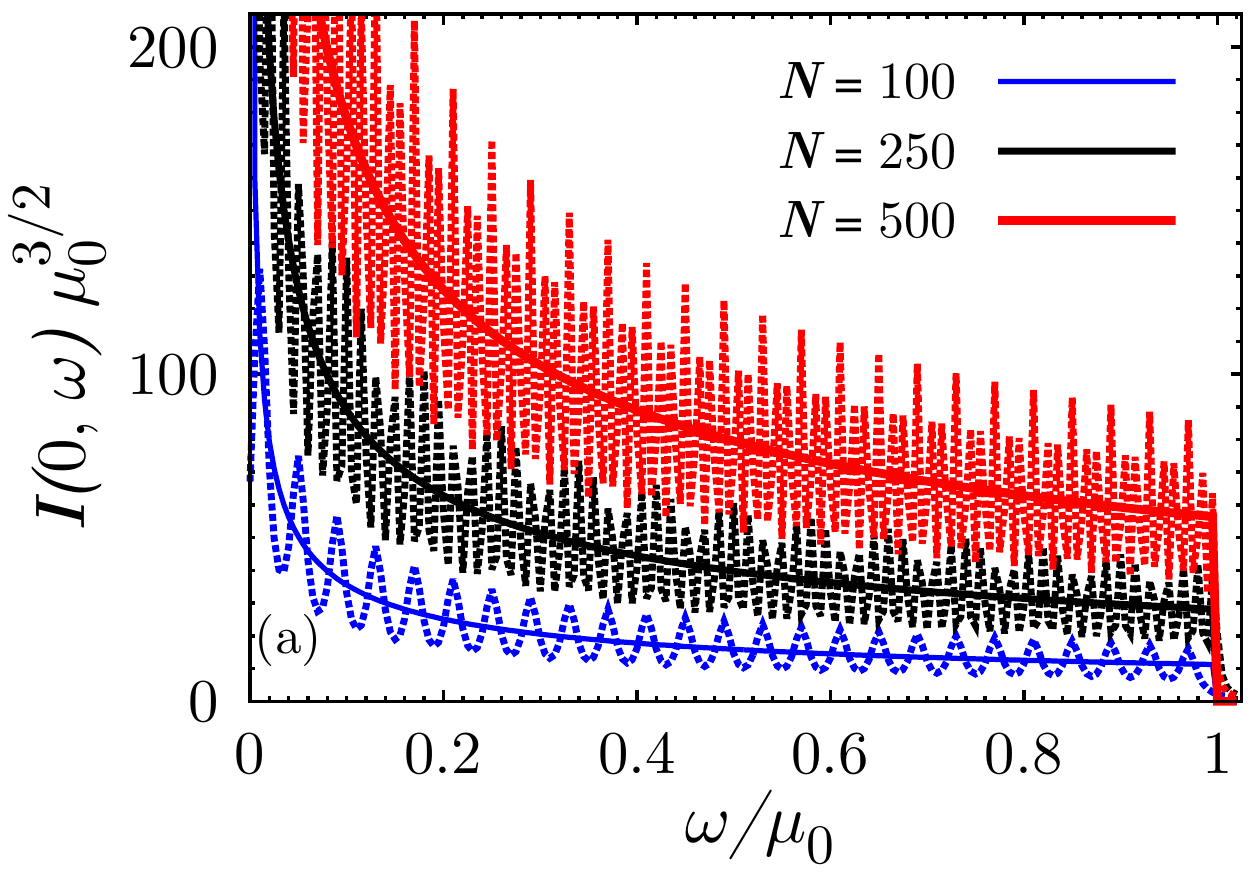}~~~~~~~~~\includegraphics[width=\myfigwidth, height=\myheight]{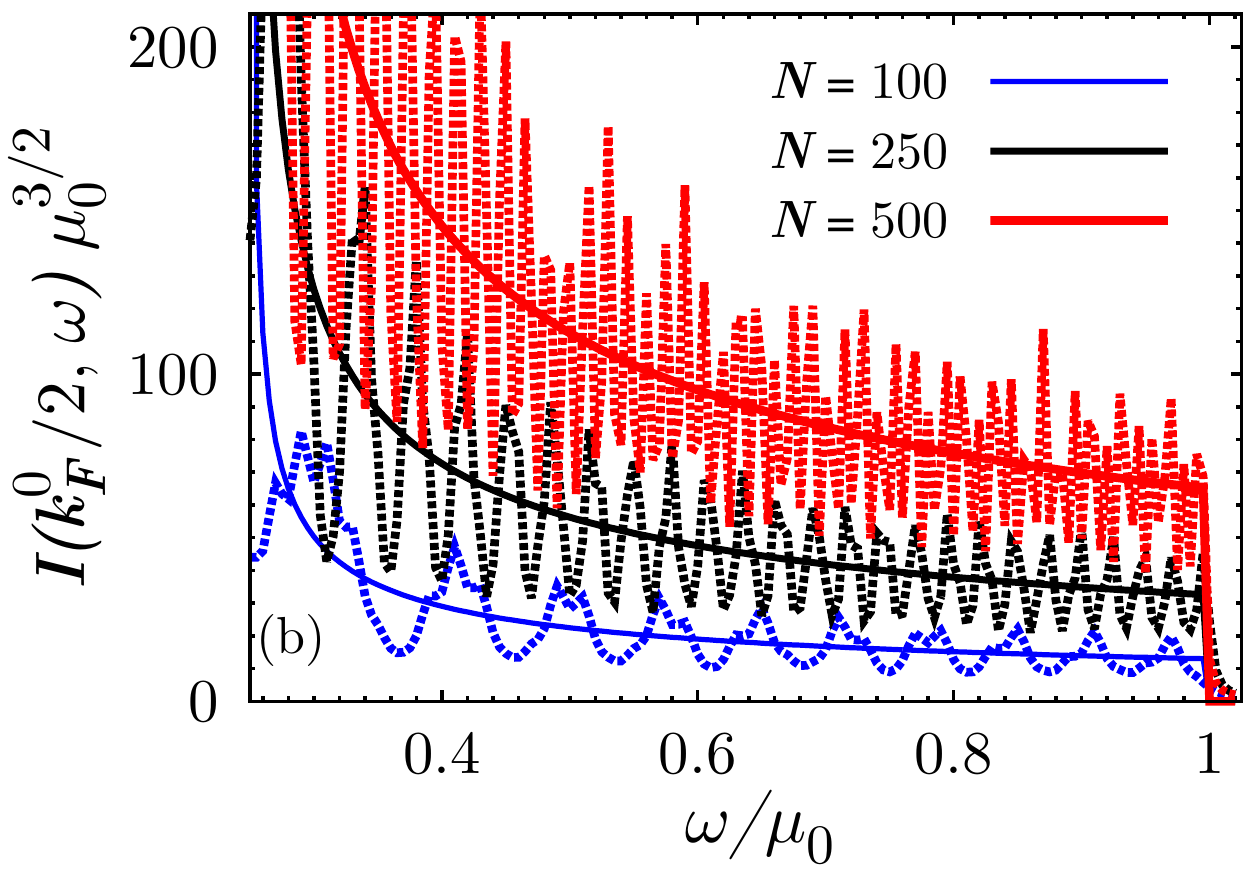}}
\caption{(Color online) The SIDF as a function of energy for a system of spin-$\half$ particles in one dimension with the broadening of the Lorentzian distribution $\frac{\eta}{\omega_0} = 0.5$ used in the exact calculation. The dotted curves denote exact results and the solid curves denote LDA results. (a) Shows the SIDF at zero momentum and  (b) Shows the SIDF at a different value of momentum $k = \frac{k_F^0}{2}$. The exact SIDF has spectral weights outside the LDA window and they have $\frac{1}{\sqrt{\omega}}$ like dependence.}
\mylabel{fig:1dk0}
\end{figure}

\begin{figure*}[!h]
\centerline{\includegraphics[width=\myfigwidth, height=\myheight]{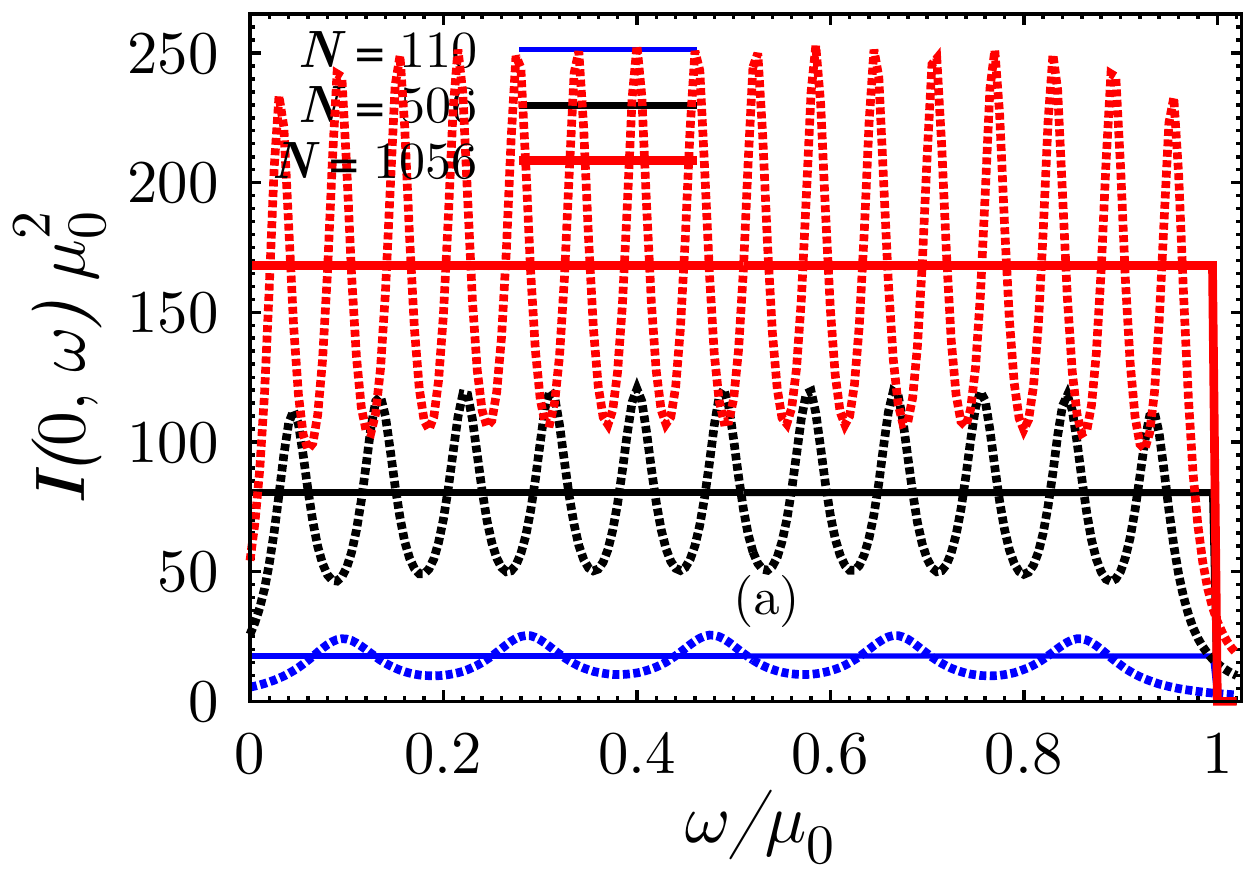}~~~~~~~~~\includegraphics[width=\myfigwidth, height=\myheight]{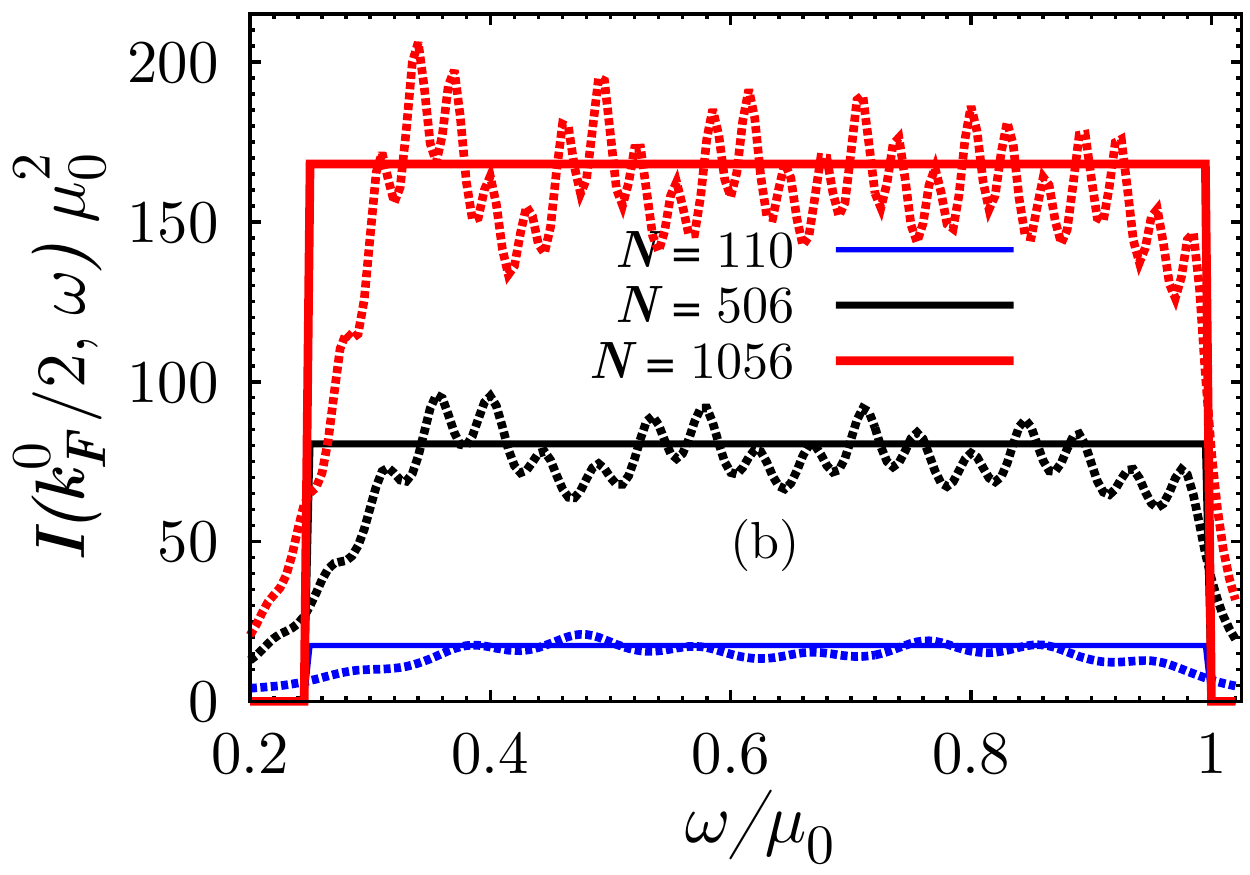}}
\caption{(Color online) For a system of spin-$\half$ fermions in two dimension, the behavior of the SIDF as a function of energy at a particular momentum with $\frac{\eta}{\omega_0} = 0.5$ used in exact calculation is shown. The dotted and the solid curves denote exact and LDA results respectively. In (a) SIDF for zero momentum and in (b) SIDF for finite momentum $k = \frac{k_F^0}{2}$ are shown. It is seen that the LDA SIDF is constant within a window but exact SIDF has weights outside that window as well. Smaller fluctuation in the exact SIDF is seen in finite momentum case.}
\mylabel{fig:2dk0}
\end{figure*}

\begin{figure*}[!t]
\centerline{\includegraphics[width=\myfigwidth, height=\myheight]{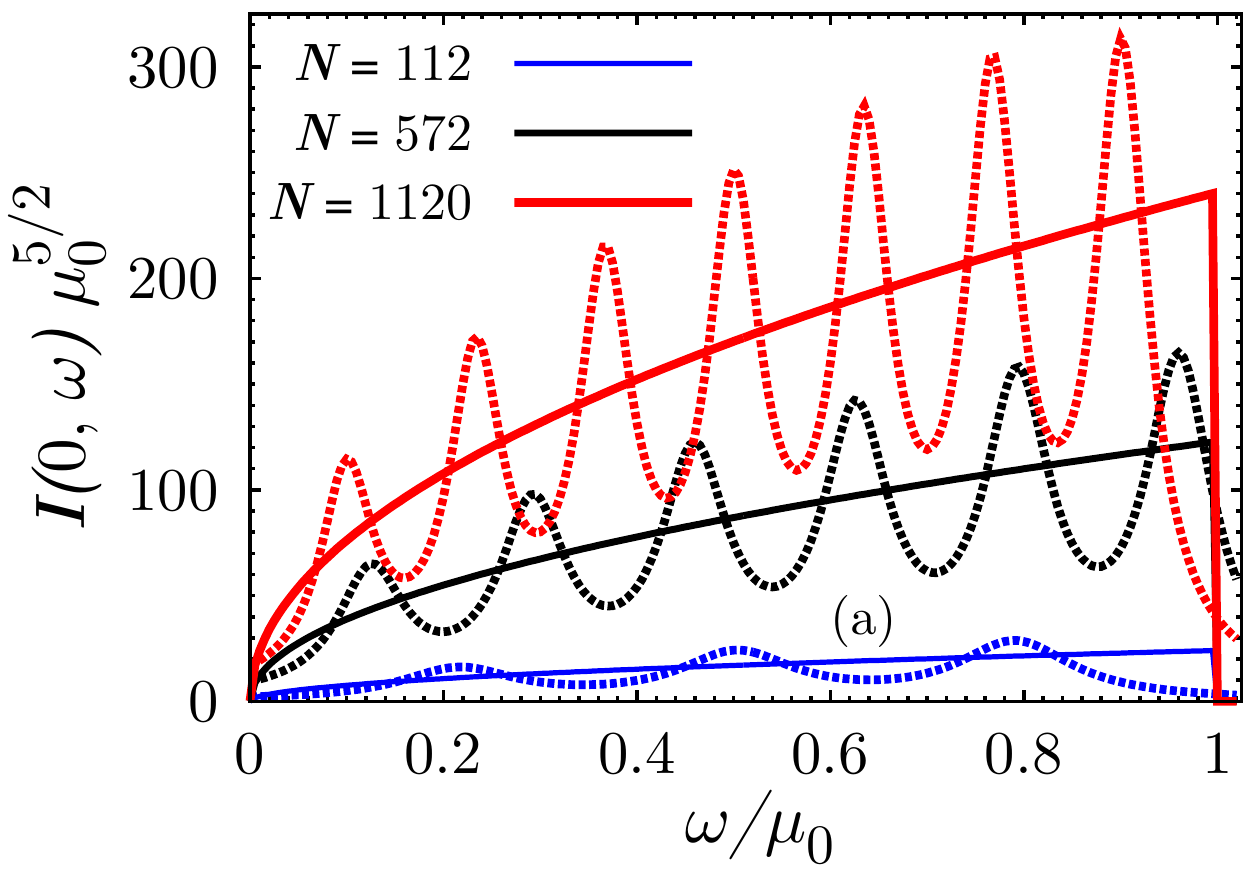}~~~~~~~~~\includegraphics[width=\myfigwidth, height=\myheight]{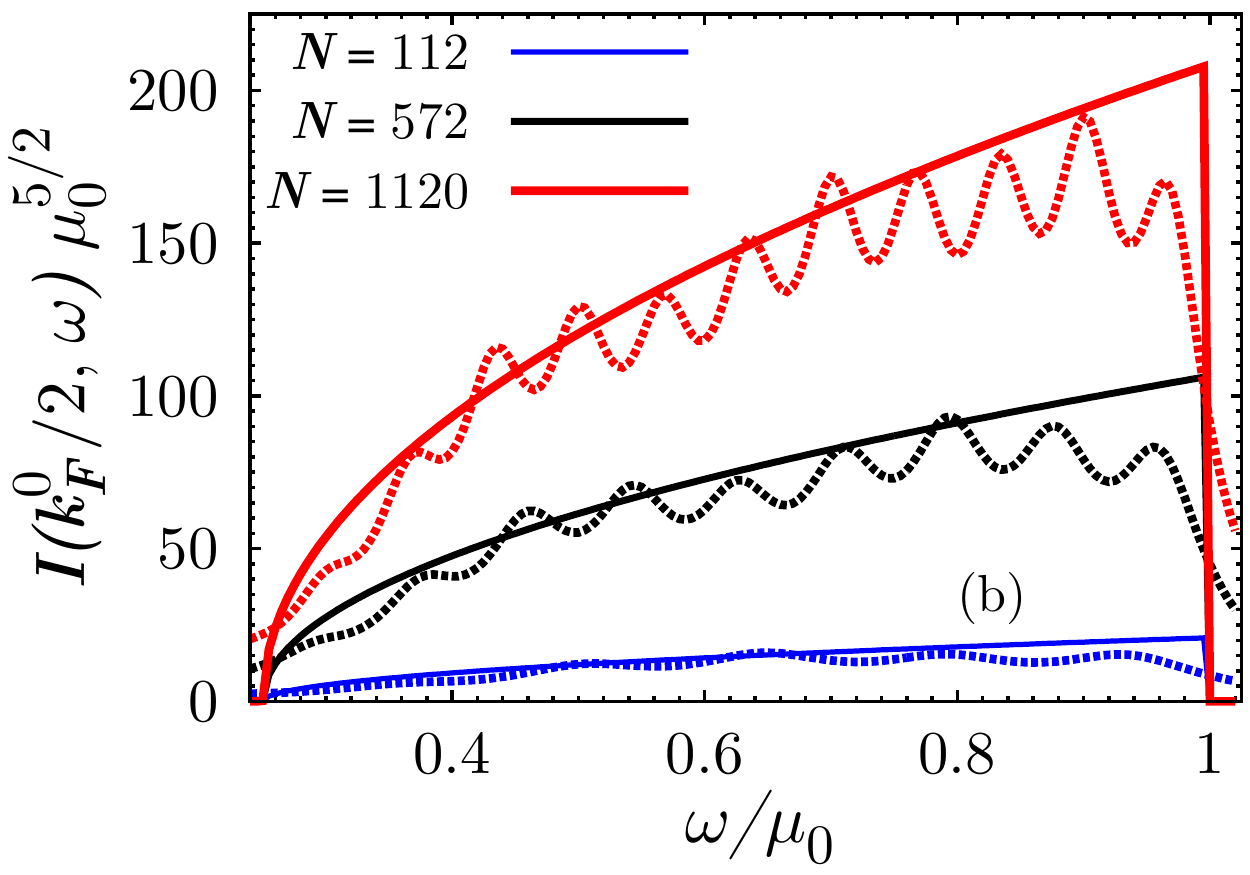}}
\caption{(Color online) The three dimensional SIDF as a function of energy at fixed momentum is shown. The value of the phenomenological parameter used in the exact calculation is $\frac{\eta}{\omega_0} = 0.5$ with exact and LDA results denoted by dotted and solid curves respectively. (a) Shows the SIDF at zero momentum and (b) Shows the SIDF at momentum $k = \frac{k_F^0}{2}$. We see that the SIDF has an overall $\sqrt{\omega}$ dependence and the exact spectra has weights outside the LDA window. The fluctuations in the exact SIDF is smaller for the case of finite momentum.}
\mylabel{fig:3dk0}
\end{figure*}

Owing to the limitation set by the trap center chemical potential and the conservation of energy, energy range of the LDA SIDF is limited to a certain window and it is non-zero only within this energy window. Although exact SIDF is large in this window but it has small spectral weight outside as well. Because of the energy time uncertainty principle, in a physical context with interaction if the excitations of the system have finite lifetime, then this leads to spectral broadening. The phenomenological broadening parameter can, therefore, be related to the finite lifetime of the quasi-particle excitations of the system. The minimum energy difference between the harmonic oscillator energy levels is $\omega_0$ and for comparing with the LDA SIDF, we have chosen the value of $\eta$ to be $\frac{\eta}{\omega_0} = \half$  in exact calculation. 
 
The comparison between the SIDFs in one dimension is shown in \fig{fig:1dk0} at different momenta $k=0$ (\fig{fig:1dk0}(a)) and $k = \frac{k_F^0}{2}$ (\fig{fig:1dk0}(b)) for $N=100$, $250$ and $500$ which correspond to highest occupied energy level $\nu_{max} = 49$, $124$ and $249$ respectively. Here, $k_F^0$ is the fermi wave vector set by the trap center chemical potential $\mu_0 = \frac{(k_F^0)^2}{2}$. We note that as the number of particles $N$ in the trap is increased, the oscillations in the exact SIDF increases. With the change in momentum, the threshold for the SIDF in LDA result is expected to change because the free particle contribution to energy is changed but there is also a high frequency cutoff in the LDA spectrum due to the trap center chemical potential. The exact SIDF has large weight within the LDA window and relatively small weight outside it; also at momentum $k = \frac{k_F^0}{2}$ but with suppressed fluctuations. We see from this figure that within corresponding window of energy the behaviors of the SIDFs for zero and finite momentum are similar.

\begin{figure*}[!t]
\centerline{\includegraphics[width=\mywidth, height=\myheightcus]{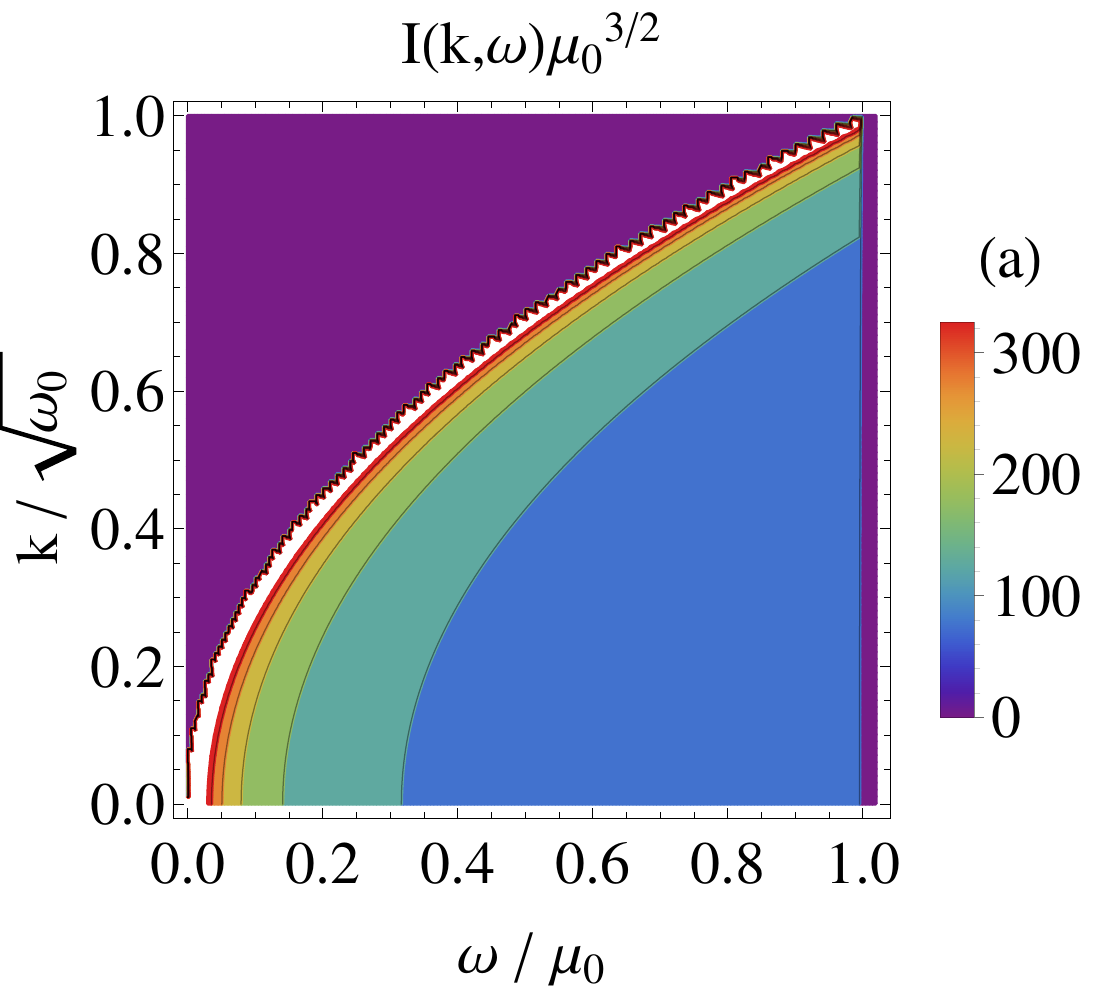}~~~\includegraphics[width=\mywidth, height=\myheightcus]{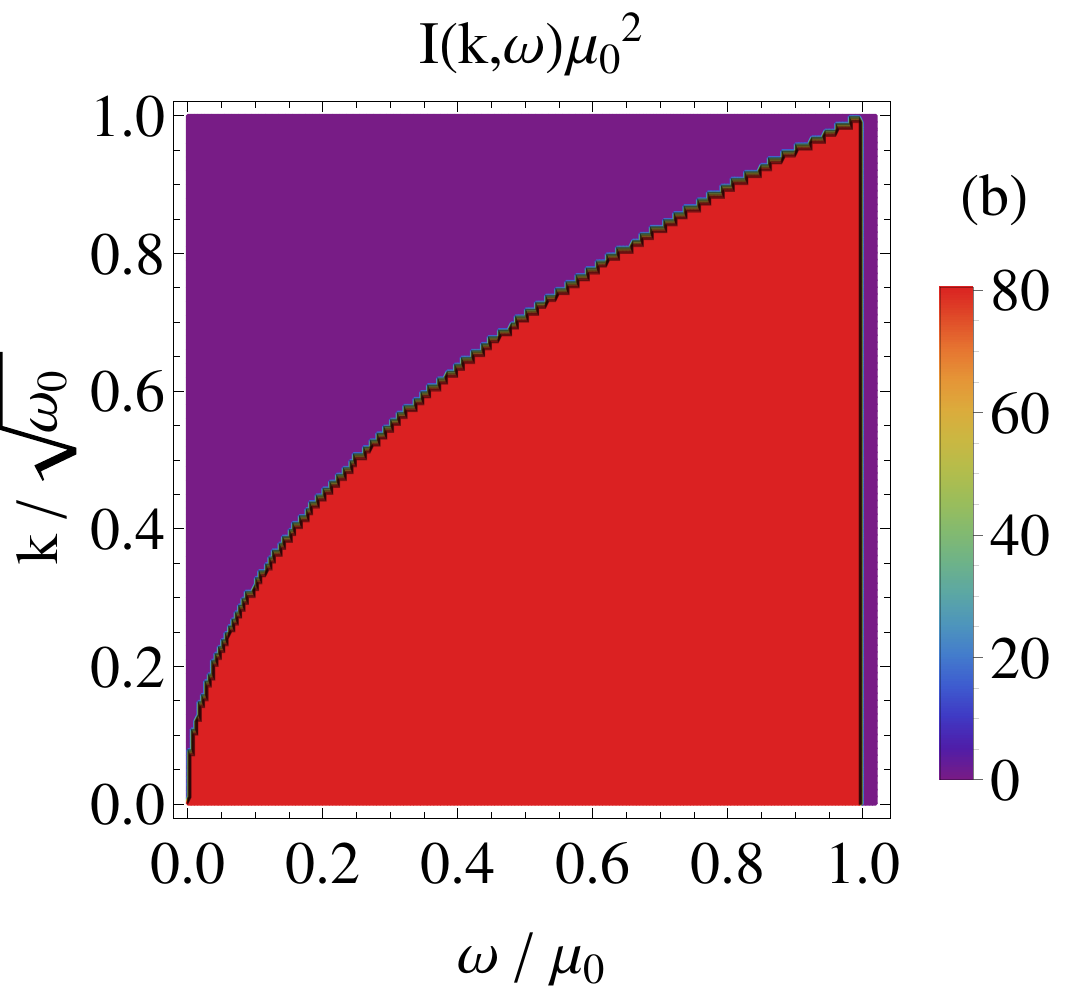}~~~\includegraphics[width=\mywidth, height=\myheightcus]{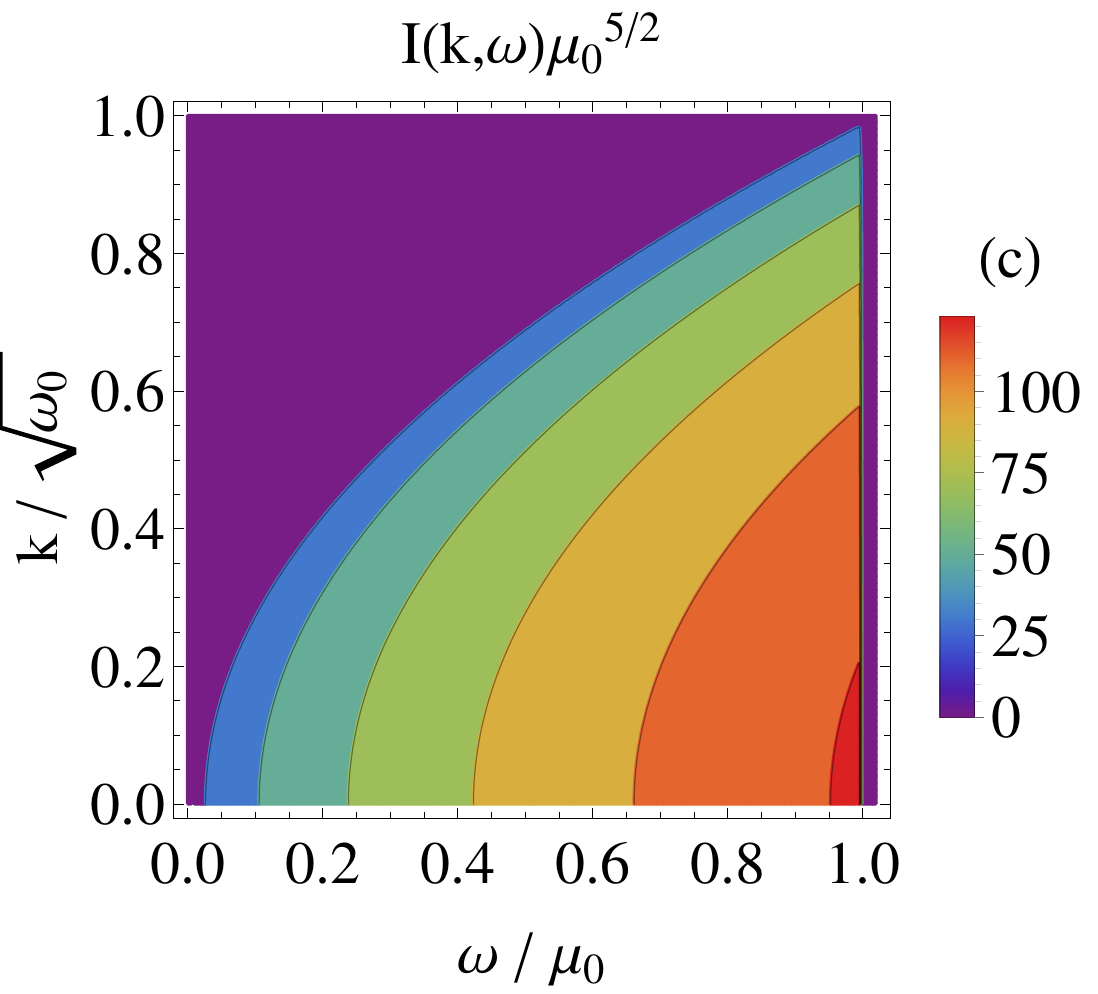}}
\caption{(Color online) Contour plot of the full SIDF spectrum as functions of both momentum and energy calculated using LDA is shown. (a), (b) and (c) show the spectra for $1$, $2$ and $3$ dimensions for number of particles $500$, $506$ and $572$ respectively. The SIDF corresponding to two dimension clearly shows that it is non-zero below a certain energy contour and is zero above it. The one dimensional SIDF for small momentum and large energies has smaller values but the three dimensional SIDF shows opposite behavior. }
\mylabel{fig:fullspec}
\end{figure*}

The expected constant behavior behavior of the SIDF in $2d$ is clearly seen from \fig{fig:2dk0}. At zero and finite momentum the comparisons of exact and LDA SIDFs are shown in \fig{fig:2dk0}(a) and \fig{fig:2dk0}(b) respectively for number particles $N = 110$, $N = 506$ and $N = 1056$ which correspond to $\nu_{max} = 9$, $21$ and $31$ respectively. It is noted that at half of the Fermi momentum, LDA gives a better description than that of the zero momentum case. As expected the LDA window becomes smaller due to the increase in the free particle energy for the finite momentum case and the comparison with the exact results becomes better for larger number of particles. 

In $3$d, $I(\bk,\omega)$ at a particular momentum $k$ as a function of $\omega$, has an overall $\sqrt{\omega}$ dependence which is as expected. This is seen form \fig{fig:3dk0} with number particles $N = 112$, $N = 572$ and $N = 1120$ which correspond to $\nu_{max} = 5$, $10$ and $13$ respectively at $k = 0$ (\fig{fig:3dk0}(a)) and $k = \frac{k_F^0}{2}$ (\fig{fig:3dk0}(b)). LDA SIDFs are in a particular window whereas exact spectra have some small spectral weights outside as well. The finite momentum SIDFs are similar to that of the zero momentum results although for finite momentum the oscillations get suppressed in the exact SIDF. In this dimension, as the number of particles in the system increases, LDA becomes better than that of 1d and 2d cases. 

The oscillations seen in the exact SIDFs for a fixed number of particles can be attributed to the discrete nature of the harmonic oscillator spectrum. The periods of these oscillations vary with changing number of particles due to the change in the trap center chemical potential. Although for comparison we have chosen a particular value of $\eta = \frac{\omega_0}{2}$, with smaller values of $\eta$ the oscillations in the exact SIDFs increase and for very small values of $\eta$, the disagreement between the exact and LDA SIDFs increases and exact SIDFs become vanishingly small. We also see step like behavior when we calculate the exact SIDF as a function of $N$ at fixed $\omega$ and $|\bk|$, due to the finite degeneracy of the harmonic oscillator levels. The same kind of behaviors were also seen earlier in \cite{Toms05,Schneider98} for this system in the thermodynamic quantities. They found that the thermodynamic quantities of this system oscillates as a function of $N$ and the continuum approximation or LDA gives better result at higher temperatures. But for very small temperature, the disagreement with the continuum approximation increases and the exact results become vanishingly small. So, the temperature there plays a similar role as the broadening parameter $\eta$ here.

In \fig{fig:fullspec} , we show the full SIDF spectrum calculated using LDA in different dimensions $1$, $2$ and $3$ (in the contour plots \fig{fig:fullspec}(a), \fig{fig:fullspec}(b) and \fig{fig:fullspec}(c) respectively) as functions of both energy and momentum. The full spectrum has the same overall features as the fixed momentum case but now the behaviors at all momenta are seen. The two dimensional spectrum clearly shows a contour separating the zero and non-zero regions. The spectra corresponding to one and three dimensions have natures opposite to each other as is expected, meaning the 1d spectrum has small spectral weights at small momenta and high energy whereas the 3d spectrum has large spectral weights in this regime.

By radio frequency photo emission spectroscopy technique, the spectral density function of a many body Fermi system is routinely measured.\cite{Jin_nat2010} Using spin polarized Fermi gas to load the atoms to a single hyperfine state, a non-interacting Fermi system can be created.\cite{Loftus2002} For RF measurements done in this kind of system of ultra dilute almost ideal Fermi gas, the SIDF of the system will have signatures as described earlier in this section. Hence, we see that although the exact dynamical response function $I(\bk,\omega)$ has many interesting features, LDA captures essentially most of them quite accurately. We believe that our results will be useful to experimentalists and stimulate further theoretical work.

\section{Summary and outlook}
\mylabel{sec:summary}
In this article, we have investigated how good is LDA in comparison to exact case in determining the dynamical response functions of an ultracold many body Fermi system. We have considered a system of ultracold dilute non-interacting spin-$\half$ Fermi gas in a harmonic trap at zero temperature. Identifying important scales in the problem, we have calculated one of the most important response functions, the SIDF of the system exactly and using the simplest version of LDA (although there are other versions of LDA which have their own domain of applicability) in any dimension. The SIDFs at a particular momentum behave as the DOS of the system without trap in the corresponding dimension. We have shown that as the dimension increases the performance of LDA gets better and increase in number of particles makes LDA better in higher dimensions. Also LDA restricts the intensity distribution function in a certain range of energies but in contrary the exact spectra always has some small weights outside the LDA window fixed by momentum and the trap center chemical potential. We have shown the full dynamical SIDF as a function of both energy and momentum calculated using LDA and expected behavior is seen. We have described the experimental significance and importance of this article. A novel future direction of this work will be to include repulsive interaction between the trapped spin-$\half$ fermions and study the behavior of the emerging Fermi liquid and its excitations in the presence of synthetic non-Abelian gauge fields.

\paragraph*{{\bf Acknowledgment :}}The author acknowledges Vijay B. Shenoy for extensive discussions. The financial support from CSIR, India via  SRF grants is thankfully acknowledged.

\bibliography{sidf}

\begin{thebibliography}{10}
\expandafter\ifx\csname url\endcsname\relax
  \def\url#1{\texttt{#1}}\fi
\expandafter\ifx\csname urlprefix\endcsname\relax\def\urlprefix{URL }\fi
\expandafter\ifx\csname href\endcsname\relax
  \def\href#1#2{#2} \def\path#1{#1}\fi

\bibitem{noble}
E.~A. Cornell, C.~E. Wieman, Rev. Mod. Phys. 74 (2002) 875--893.

\bibitem{Greiner_nat2002}
M.~Greiner, O.~Mandel, T.~Esslinger, T.~W. Hansch, I.~Bloch, Nature 415 (2002)
  39--44.

\bibitem{Regal2006}
C.~A. Regal, D.~S. Jin, Adv. Atom. Mol. Opt. Phys. 54 (2006) 1--79.

\bibitem{Lin_nat2009}
Y.-J. Lin, R.~L. Compton, K.~J. Garcia, J.~V. Porto, I.~B. Spielman, Nature 462
  (2009) 628--632.

\bibitem{Philips2009}
Y.-J. Lin, R.~L. Compton, A.~R. Perry, W.~D. Phillips, J.~V. Porto, I.~B.
  Spielman, Phys. Rev. Lett. 102 (2009) 130401.

\bibitem{Dalibard2011}
J.~Dalibard, F.~Gerbier, G.~Juzeli\ifmmode~\bar{u}\else \={u}\fi{}nas,
  P.~\"Ohberg, Rev. Mod. Phys. 83 (2011) 1523--1543.

\bibitem{Sudeep2011}
S.~K. Ghosh, J.~P. Vyasanakere, V.~B. Shenoy, Phys. Rev. A 84 (2011) 053629.

\bibitem{Jayantha_twobody}
J.~P. Vyasanakere, V.~B. Shenoy, Phys. Rev. B 83 (2011) 094515.

\bibitem{Hu2011}
H.~Hu, L.~Jiang, X.-J. Liu, H.~Pu, Phys. Rev. Lett. 107 (2011) 195304.

\bibitem{Jayantha2011}
J.~P. Vyasanakere, S.~Zhang, V.~B. Shenoy, Phys. Rev. B 84 (2011) 014512.

\bibitem{Wang2012}
P.~Wang, Z.-Q. Yu, Z.~Fu, J.~Miao, L.~Huang, S.~Chai, H.~Zhai, J.~Zhang, Phys.
  Rev. Lett. 109 (2012) 095301.

\bibitem{Cheuk2012}
L.~W. Cheuk, A.~T. Sommer, Z.~Hadzibabic, T.~Yefsah, W.~S. Bakr, M.~W.
  Zwierlein, Phys. Rev. Lett. 109 (2012) 095302.

\bibitem{Comin2013}
R.~Comin, A.~Damascelli, arXiv preprint arXiv:1303.1438.

\bibitem{Andrea2003}
A.~Damascelli, Z.~Hussain, Z.-X. Shen, Rev. Mod. Phys. 75 (2003) 473--541.

\bibitem{Dao2007}
T.~L. Dao, A.~Georges, J.~Dalibard, C.~Salomon, I.~Carusotto, Phys. Rev. Lett.
  98 (2007) 240402.

\bibitem{Jin_nat2008}
J.~T. Stewart, J.~P. Gaebler, D.~S. Jin, Nature 454 (2008) 744--747.

\bibitem{Jin_nat2010}
J.~P. Gaebler, J.~T. Stewart, T.~E. Drake, D.~S. Jin, A.~Perali, P.~Pieri,
  G.~C. Strinati, Nature Physics 6 (2010) 569--573.

\bibitem{Brend2011}
B.~Fr\"ohlich, M.~Feld, E.~Vogt, M.~Koschorreck, W.~Zwerger, M.~K\"ohl, Phys.
  Rev. Lett. 106 (2011) 105301.

\bibitem{Ville2012}
V.~Pietil\"a, Phys. Rev. A 86 (2012) 023608.

\bibitem{Kohn1985}
E.~K.~U. Gross, W.~Kohn, Phys. Rev. Lett. 55 (1985) 2850--2852.

\bibitem{Yabana1999}
K.~Yabana, G.~F. Bertsch, Int. Journal of Quant. Chem. 75 (1999) 55--66.

\bibitem{Stringari04}
L.~Pezz\`e, L.~Pitaevskii, A.~Smerzi, S.~Stringari, G.~Modugno, E.~de~Mirandes,
  F.~Ferlaino, H.~Ott, G.~Roati, M.~Inguscio, Phys. Rev. Lett. 93 (2004)
  120401.

\bibitem{Mueller04}
E.~J. Mueller, Phys. Rev. Lett. 93 (2004) 190404.

\bibitem{Gleisberg00}
F.~Gleisberg, W.~Wonneberger, U.~Schl\"oder, C.~Zimmermann, Phys. Rev. A 62
  (2000) 063602.

\bibitem{Vignolo00}
P.~Vignolo, A.~Minguzzi, M.~P. Tosi, Phys. Rev. Lett. 85 (2000) 2850--2853.

\bibitem{Butts97}
D.~A. Butts, D.~S. Rokhsar, Phys. Rev. A 55 (1997) 4346--4350.

\bibitem{Toms05}
D.~J. Toms, Annals of Physics 320 (2005) 487--520.

\bibitem{Schneider98}
J.~Schneider, H.~Wallis, Phys. Rev. A 57 (1998) 1253--1259.

\bibitem{Loftus2002}
T.~Loftus, C.~A. Regal, C.~Ticknor, J.~L. Bohn, D.~S. Jin, Phys. Rev. Lett. 88
  (2002) 173201.

\end{thebibliography}

\end{document}